\documentclass[aps,prb,twocolumn,floatfix,superscriptaddress]{revtex4-2}
\usepackage{psfrag,epsfig,amsfonts,amssymb,amsmath,wasysym,bm}
\usepackage{bbold}
\usepackage{color}

\usepackage[inline]{enumitem} 
\usepackage{hyperref}

\newcommand{\<}{\langle}	
\renewcommand{\>}{\rangle}	
\newcommand{\ket}[1]{\lvert #1 \rangle} 	
\newcommand{\bra}[1]{\langle #1 \rvert}	
\newcommand{\ketN}[1]{\lvert #1 \rangle_{\! 0}} 	
\newcommand{\braN}[1]{{_0\!}\langle #1 \rvert}	
\newcommand{\mc}{\mathcal}	
 
\renewcommand{\Im}{\operatorname{Im}}

\newcommand{\e}{\mathrm{e}}		
\newcommand{\I}{\mathrm{i}}		

\newcommand{\id}{\mathbb 1}
\newcommand{\RR}{{\mathbb R}}

\newcommand{\NN}{{\mathbb N}}

\newcommand{\tr}{\operatorname{tr}}

\newcommand{\Autav}{\bar A_0}
\newcommand{\Auth}{A_{0,\mathrm{th}}}
\newcommand{\Ainf}{A_{\infty}}

\newcommand{\varV}{\tilde{v}}
\newcommand{\varVFT}{v}

\providecommand{\av}[1]{\mathbb{E}[#1]}

\newcommand{\lav}[2]{[#1]_{#2}}

\renewcommand{\d}{\mathrm{d}} 
\newcommand{\lmat}{\left( \begin{matrix}}	
\newcommand{\rmat}{\end{matrix} \right)}	

\begin{document}

\title{Random-matrix approach to time-dependent forcing in many-body quantum systems}

\author{Lennart Dabelow}
\affiliation{School of Mathematical Sciences, Queen Mary University of London, London E1 4NS, UK}

\author{Peter Reimann}
\affiliation{Faculty of Physics, 
Bielefeld University, 
33615 Bielefeld, Germany}
\date{\today}

\begin{abstract}

Changing some of 
its parameters over time is a paradigmatic way 
of driving an otherwise isolated many-body quantum system out of 
equilibrium,
and a vital ingredient for building quantum computers and simulators.
Here, we 
further develop a recently proposed
nonlinear response theory
which is based on typicality 
and random-matrix methods, 
and which is
applicable to a wide variety of 
such parametrically perturbed systems
in and out of equilibrium:
We derive analytical approximations of the characteristic response function for the two limiting cases
of fast driving and of strong and short-ranged-in-energy driving.
Furthermore, we work out implications and predictions for common applications,
including finite-time quenches and time-dependent forcing that breaks conservation laws of the underlying undriven system.
Finally,
we verify all predictions 
by
numerical examples
and discuss the theory's scope and limitations.
\end{abstract}

\maketitle

\section{Introduction}

It is a fundamental problem to deduce how macroscopic systems respond to 
time-dependent variations of some intrinsic properties or external control 
parameters directly from the well-established laws of quantum mechanics 
which govern
their microscopic constituents.
Moreover, precisely targeted manipulations of complex quantum systems by time-dependent forcing have become technically feasible and increasingly widespread in recent years,
e.g., in the controlled setting of cold-atom \cite{
gre02, man03, kin06,
blo08, blo12, bla12, geo14, lan15, bor17, rub20, ued20}
or polarization-echo 
\cite{geo14, gar17, wei18, wei19, nik20, pen21, bea21} experiments.
Not least, such manipulations are key for operating general-purpose quantum computers and simulators \cite{nie10, blo12, bla12, geo14}.
Finally,
periodic driving in particular has been suggested as a way to create ``time crystals'' and various meta-materials with intriguing topological properties \cite{
lin11,
aid13, miy13,
gru14, 
hol16, cho17, els17, moe17, oka19, mac20,
win20,
wei21, 
vie21, min21}.
For a comprehensive understanding and efficient control of those diverse applications, 
it is of great interest to describe the generic effects of time-dependent manipulations on large quantum systems in a reasonably general setting.

A theoretical framework to obtain such general insights was brought forward in Ref.~\cite{dab24}
using typicality and random-matrix methods.
Instead of looking at the response of a system to one particular type of time-dependent forcing,
the idea is to consider an ensemble of different, but similar drivings applied to the system.
By demonstrating that the observable response is practically indistinguishable for nearly all members of the ensemble,
the average behavior becomes a prediction for the response of an individual system.
In spirit, this idea is thus similar to using averages over, say, a microcanonical ensemble to predict equilibrium properties of a concrete physical system.
Adopting this approach,
a central result of Ref.~\cite{dab24} was an analytical theory describing the observable response 
of many-body quantum systems,
both in and out of equilibrium, to time-dependent driving.

Here we extend this theory in several ways.
As reviewed in Sec.~\ref{sec:Theory},
the characteristic response function is obtained as the solution of a nonlinear integro-differential equation,
which must usually be solved numerically.
In Sec.~\ref{sec:ResProf}, we present approximate analytical solutions of this equation in two limiting cases,
(i) when the driving is sufficiently strong (large amplitude)
and short-ranged-in-energy, or (ii) when it is
sufficiently fast (short characteristic time scale).
We also map out the validity regime of these approximations as well as of the general solution with regard to properties of the system and driving.
In Secs.~\ref{sec:FiniteQuench}--\ref{sec:DoublePretherm},
we then explore the consequences and predictions of the theory for commonly encountered application settings:
finite-time quenches in Sec.~\ref{sec:FiniteQuench},
patternless (pseudorandom) driving in Sec.~\ref{sec:PseudrandomDriving},
and breaking of conservation laws in Sec.~\ref{sec:DoublePretherm},
which leads to a ``double prethermalization'' effect, in particular.
We conclude with a few remarks on the theory's general applicability in Sec.~\ref{sec:Conclusions}.

\section{Typical nonlinear response}
\label{sec:Theory}

In this section, we recall the prediction for the observable response of many-body quantum systems to time-dependent driving,
first derived in Ref.~\cite{dab24},
along with the principal assumptions.

\subsection{Setup}
\label{sec:Theory:Setup}

We consider the dynamics of many-body quantum systems, described in terms of the time-dependent expectation values
\begin{equation}
\label{eq:TimeEvo}
	\< A \>_{\!\rho(t)} := 
\tr\{ \rho(t) A \}
\end{equation}
of experimentally realistic, physical observables $A$.
Here $\rho(t) := \mc U(t) \rho(0) \mc U^\dagger(t)$ is the state at time $t$ of the system that is prepared in some (pure or mixed) initial state $\rho(0)$ at $t = 0$,
and the propagator $\mc U(t)$ satisfies $\frac{\d}{\d t} \mc U(t) = -\I H(t) \mc U(t)$ 
and $\mc U(0) = \id$ (identity operator).
The corresponding Hamiltonians are of the general form
\begin{equation}
\label{eq:H}
	H(t) := H_0 + f(t) \, V \,,
\end{equation}
where $H_0$ is a time-independent 
Hamiltonian of some unperturbed reference system
and the (constant) {\em driving operator} $V$ corresponds to a perturbation,
whose coupling strength varies in time as described by the (scalar) {\em driving protocol} $f(t)$.
In Ref.~\cite{dab24}, the focus was on periodic protocols, but we will admit largely arbitrary functions in the following.

The time-independent (unperturbed)
reference system with Hamiltonian $H_0$ 
is assumed to
exhibit a well-defined macroscopic energy.
This 
means
that there exists a macroscopically small energy interval $\Delta$
containing all the significantly occupied 
levels of the unperturbed system in the initial state $\rho(0)$.
In particular, 
denoting the energy levels of $H_0$ by $E_\mu$,
we take it for granted
that the (suitably coarse-grained) density of states
$D(E) := \sum_\mu \delta(E - E_\mu)$
can be approximated by a constant $D(E) \approx D_0$ throughout $\Delta$.
Note that initial conditions which mainly populate levels
close to the edges of the spectrum are therefore tacitly excluded.
On the other hand,
the initial state $\rho(0)$ is not required
to be a thermal equilibrium ensemble or steady state of the unperturbed
system $H_0$.

Denoting the
eigenstates of $H_0$
by $\ketN{\mu}$ such that $H_0 \ketN{\mu} = E_\mu \ketN{\mu}$, the 
{\em thermal equilibrium expectation value}
of the observable $A$ associated with this microcanonical energy window $\Delta$ is
\begin{equation}
\label{eq:Auth}
	\Auth := \frac{1}{N} \sum_{\mu : E_\mu \in \Delta} \braN{\mu} A \ketN{\mu} \,,
\end{equation}
where $N$ is the number of levels in $\Delta$.
Due to the extremely high level density of generic many-body systems, $N$ is a huge number,
usually growing exponentially with the system's degrees of freedom \cite{lan70, gol10a}.

The observable dynamics of this unperturbed system,
\begin{equation}
\label{eq:TimeEvoUndriven}
	\< A \>_{\!\rho_0(t) }:= \tr\{ \rho_0(t) A \}
\end{equation}
with $\rho_0(t) := \e^{-\I H_0 t} \rho(0) \e^{\I H_0 t}$,
is assumed to be known (e.g.\ from measurements or because $H_0$ is particularly simple)
and is one ingredient of the theory.

Our
key 
assumption regarding
the 
driving operator
$V$ is formulated in terms of its matrix elements $V_{\mu\nu} := \braN{\mu} V \ketN{\nu}$ 
in the eigenbasis of $H_0$.
We define the associated
\emph{perturbation profile} as
\begin{equation}
\label{eq:VarV}
	\varV(E) := \lav{ \lvert V_{\mu\nu} \rvert^2 }{E} \,,
\end{equation}
where $\lav{ \,\cdots }{E}$ denotes a local average over matrix elements with $\lvert E_\mu - E_\nu \rvert \approx E$.
The principal assumption about $V$ is that $\varV(E)$ is a well-defined function,
i.e.,
within the relevant energy window $\Delta$, the average on the right-hand side 
of~\eqref{eq:VarV} depends only on (the absolute value of) the difference $E_\mu - E_\nu$ of the coupled levels
(and not on $E_\mu$ and $E_\nu$ separately).
The validity of this assumption for physically reasonable perturbations is supported by 
general semiclassical arguments \cite{fei89, fyo96}, analytical studies of lattice systems 
\cite{ara16, oli18} as well as numerous  concrete examples 
\cite{gen12, beu15, kon15, bor16, jan19, dab20modification, ric20}.
It is also akin to the typical matrix structure conjectured by the 
eigenstate thermalization hypothesis (ETH) \cite{deu91, sre94, rig08},
where the matrix elements $V_{\mu\nu}$ are viewed as
``pseudo-random variables'', whose statistics in general depend 
on the difference of the corresponding energies $E_\mu$ 
and $E_\nu$ as well as on their sum.
The latter, however, is (approximately) negligible within our
present narrow energy window $\Delta$ with (approximately)
constant level density (see above). 
Similarly, the mean value of the diagonal ETH-matrix 
elements will be (approximately) constant within $\Delta$
and can be set to zero without loss of generality.
We also remark that our assumption of a well-defined perturbation 
profile in (\ref{eq:VarV}) still admits a very large class of
possibly banded and/or sparse matrices $V_{\mu\nu}$
(see also Sec. \ref{sec:Theory:Result} below).

A quantity that will turn out to be of particular relevance for our main result is
the perturbation profile's Fourier transform
\begin{equation}
\label{eq:VarVFT}
	\varVFT(t) := \int \d E \, D_0 \, \e^{\I E t} \, \varV(E) \,.
\end{equation}

The driving protocol $f(t)$ is largely arbitrary,
except that the 
time-dependent perturbations $f(t) V$ in (\ref{eq:H})
should not become overly strong compared to $H_0$, 
so that establishing a connection between the unperturbed and 
perturbed systems remains reasonable.
In practice, this essentially means that $H(t)$ and $H_0$ should exhibit 
similar thermodynamic properties for any fixed $t$.
Notably, $f(t)$ need \emph{not} be periodic.
The key characteristics of the driving protocol involve its first and second 
integrals, 
\begin{equation}
\label{eq:F}
F_1(t) := \int_0^t \d s \, f(s) 
\quad \text{and} \quad 
F_2(t) := \int_0^t \d s \, F_1(s)
\,.
\end{equation}
From these we define the two auxiliary functions
\begin{equation}
\label{eq:phi}
	\varphi_1(t) := \left[ \frac{F_1(t)}{t} \right]^2
	\ \text{and} \ \ 
	\varphi_2(t) := \left[ \frac{F_2(t)}{t} - \frac{F_1(t)}{2} \right]^2 ,
\end{equation}
whose magnitude will turn out below to capture the effective perturbation strength.

\subsection{Predicted response}
\label{sec:Theory:Result}

Based on these assumptions,
it was established in Ref.~\cite{dab24}
that the overwhelming majority of systems with the same $H_0$, $\varV(E)$ and $f(t)$
exhibit the following \emph{typical}, nonlinear response behavior:
\begin{equation}
\label{eq:TypTimeEvo}
	\< A \>_{\!\rho(t)}
		= \Auth + \lvert \gamma(t, t) \rvert^2 \left[ \< A \>_{\!\rho_0(t)} - \Auth \right] .
\end{equation}
The details of the response are described by the
{\em response function}
$\gamma(t, t')$, evaluated at $t' = t$ in~\eqref{eq:TypTimeEvo},
which
obeys the
nonlinear integro-differential equation
\begin{equation}
\label{eq:ResProfEq}
\begin{aligned}
	&\frac{\partial \gamma(t, t')}{\partial t} \\
		&= - \int_0^t \!\! \d s \, \gamma(t-s, t') \gamma(s, t') 
		\! \left[ \varphi_1(t') - \varphi_2(t') \frac{\partial^2}{\partial s^2} \right] \! \varVFT(s)
\end{aligned}
\end{equation}
with initial condition
$\gamma(0, t') = 1$ for all $t'$.

Formally, this result is derived for a random-matrix model that implements the key property~\eqref{eq:VarV} of the driving operator in an ergodic sense:
Instead of the particular $V$ in the setup of interest,
we consider an ensemble of operators that share the property~\eqref{eq:VarV} on average.
More specifically, $V_{\mu\nu}$ are independent (apart from $V_{\mu\nu}^* = V_{\nu\mu}$), unbiased random variables with variance $\varV(E_\mu - E_\nu)$ and an otherwise largely arbitrary distribution
thanks to a generalized central limit theorem.
The main insight is that nearly all matrices generated in this way exhibit the same typical time evolution~\eqref{eq:TypTimeEvo},
i.e., sample-to-sample fluctuations are exponentially suppressed in the number of degrees of freedom.
In the absense of specific counterarguments,
we thus expect that the effects of the 
original (non-random) perturbation $V$ of actual interest are captured by this 
prediction~\eqref{eq:TypTimeEvo} as well;
see also the discussions in Sec.~\ref{sec:Conclusions} and Ref.~\cite{dab24}.
Note that arguments of this type are at the very heart of 
random-matrix theory and their validity is 
commonly taken for granted.
 
For completeness, we recall a few key steps of the derivation in Appendix~\ref{app:Derivation}.
In particular, it
involves a Magnus expansion \cite{bla09},
truncated at second order,
which
restricts the applicability of the result~\eqref{eq:TypTimeEvo} to the initial transient dynamics up to a time scale $t_*$.
This time scale becomes larger as
the characteristic time scale $T$ of $f(t)$
(e.g., the period for periodic driving
or the duration of a finite-time quench)
becomes smaller.
A natural limit is set by the scale at which the Magnus expansion breaks down.
Since this breakdown has been related to the onset of heating in perpetually 
(periodically) driven systems \cite{dal13, dal14, ish18},
the result~\eqref{eq:TypTimeEvo} does not capture this stage of 
the dynamics anymore.
This is also reflected in the fact that the baseline
equilibrium
value occurring in~\eqref{eq:TypTimeEvo} is
the thermal equilibrium value $\Auth$ associated 
with the initially occupied energy window $\Delta$ 
of the unperturbed reference system $H_0$, and not, for instance, the expectation value
\begin{equation}
	\Ainf := \tr \{\rho_\infty A\}
\end{equation}
in the infinite-temperature state
$\rho_\infty := \id/\tr\{\id\}$, as expected generically at very late times 
for periodically driven many-body systems with a bounded spectrum
\cite{dal14, laz14equilibrium,
pon15manybody,
ish18, mal19heating}.
Yet it has been shown under rather general circumstances 
\cite{aba15, mor16, aba17rigorous, aba17effective} 
(and verified experimentally \cite{rub20, pen21, bea21})
that heating is suppressed exponentially in the driving frequency,
meaning that $t_*$ can still comprise several multiples of $T$
if the latter is sufficiently small.
This is explored in more detail in Sec.~\ref{sec:ResProf}.

Provided that the setup of interest is a typical member of the considered 
model classes (see above Eq.~\eqref{eq:TypTimeEvo}),
the combination of Eqs.~\eqref{eq:TypTimeEvo} and~\eqref{eq:ResProfEq}
thus represents an analytical prediction for the observable expectation values 
of the driven system at arbitrary times during the initial (transient)  phase.

\section{Analytical approximations}
\label{sec:ResProf}

The key 
player in
the theoretical prediction~\eqref{eq:TypTimeEvo} is the
response function $\gamma(t, t')$,
which connects the undriven dynamics to the typical behavior of the driven system
and satisfies Eq.~\eqref{eq:ResProfEq}.
In this section, we present analytical solutions of Eq.~\eqref{eq:ResProfEq} in two limiting cases.
While it is straightforward in principle to compute $\gamma(t, t')$ for any given perturbation profile $\varV(E)$ 
and driving protocol $f(t)$ by directly integrating Eq.~\eqref{eq:ResProfEq} numerically,
the analytical approximations in closed form provide further and more direct insights into the structure of our prediction~\eqref{eq:TypTimeEvo}.

\subsection{Strong and short-ranged-in-energy driving}
\label{sec:ResProf:StrongDriving}

The first approximation targets the 
realm
of strong and short-ranged-in-energy driving,
meaning large amplitudes of $f(t)$ in Eq.~\eqref{eq:H}
and narrow perturbation profiles (fast decay of $\varV(E)$ with $\lvert E \rvert$) 
in Eq.~\eqref{eq:VarV}.
The expected regime of applicability will be determined precisely below by a self-consistency requirement [see the discussion after Eq.~\eqref{eq:ResProf:StrongDriving:Condition}].
Our starting point is the following representation for $\gamma(t, t')$,
\begin{equation}
\label{eq:ResProfFromG}
	\gamma(t, t') = \frac{1}{\pi} \lim_{\eta\to 0+} \int \d E \, \e^{\I E t} \, \Im G(E - \I\eta, t') \,,
\end{equation}
involving
the Fourier transform of the ensemble-averaged resolvent $G(z - H_0, t') := \av{(z - H^{(t')})^{-1}}$ 
of a family of auxiliary Hamiltonians 
\begin{equation}
H^{(t')} := H_0 + \frac{F_1(t')}{t'} V + [\frac{F_2(t')}{t'} - \frac{F_1(t')}{2}] \I [V, H_0]
\ .
\label{neu1}
\end{equation}
The ensemble-averaged resolvent $G(z, t')$ satisfies an integral equation of the form
\begin{widetext}
\begin{equation}
\label{eq:GAuxEq}
	G(z, t') \left\{ z - \int \d E \, D_0 \, G(z-E, t') \left[ \varphi_1(t') - E^2 \varphi_2(t') \right] \varV(E) \right\} = 1 \,.
\end{equation}
\end{widetext}
This connection was established in Ref.~\cite{dab24}, see Eqs.~(23) and~(24) therein;
see also Appendix~\ref{app:Derivation}
for a few more details on the context in which it emerges.

Let us assume that the perturbation profile $\varV(E)$ and also $E^2 \varV(E)$ decay much faster as $\lvert E \rvert \to \infty$ than the typical scale on which the resolvent $G(z - E, t')$ varies for any relevant, fixed $z$ and $t'$.
Since $\varV(-E) = \varV(E)$ [cf.\ Eq.~\eqref{eq:VarV}],
we can then approximate $G(z - E, t') \approx G(z, t')$ in the integrand of~\eqref{eq:GAuxEq}.
Consequently, the latter equation becomes
\begin{equation}
\label{eq:IntEqG:StrongDriving}
	[ r(t') \, G(z, t') ]^2 / 4 - z \, G(z, t') + 1 = 0 \,,
\end{equation}
with
\begin{equation}
\label{eq:ResProf:StrongDrivingScale}
	 r(t) := \sqrt{ 4 \varV(0) D_0 \left[ \Sigma_0 \varphi_1(t) + \Sigma_2 \varphi_2(t) \right] } \, ,
\end{equation}
where the functions $\varphi_1(t)$ and $\varphi_2(t)$ were defined in~\eqref{eq:phi} and
\begin{equation}
\label{eq:VarV:Moments}
	\Sigma_n := \frac{1}{\varV(0)} \int \d E \, E^n \, \varV(E) \,.
\end{equation}

The solution of the algebraic equation~\eqref{eq:IntEqG:StrongDriving} for $G(z, t')$ is
\begin{equation}
\label{eq:ResProf:StrongDriving:GAux}
	G(z, t') = \frac{2}{r(t')^2} \left[ z - \I \operatorname{sgn}(\Im z) \sqrt{r(t')^2 - z^2} \right] ,
\end{equation}
taking into account that $\Im G(z, t')$ and $\Im z$ must have opposite signs \cite{dab24}.

Under the above assumption of a sufficiently 
narrow (fast decaying) perturbation profile $\varV(E)$,
the typical scale on which $G(z, t')$ varies with the energy argument $z$ for fixed auxiliary time $t'$
is thus $r(t')$.
Self-consistency of the initial assumption regarding
the relation between the scales of $G(z, t')$ and $\varV(E)$  therefore demands that
\begin{equation}
\label{eq:ResProf:StrongDriving:Condition}
	r(t') \gg \Sigma_0
\end{equation}
since $\Sigma_0$ is a natural measure of the width of $\varV(E)$ [e.g., $\Sigma_0 = 2 \Delta_v$ for the exponential profile from Eq.~\eqref{eq:VarV:Exp} below].
Due to the parametric dependence of the condition~\eqref{eq:ResProf:StrongDriving:Condition} on the auxiliary time $t'$
[which is set to $t' = t$ in the main result~\eqref{eq:TypTimeEvo}],
the quality of the present approximation will not be uniform in time.

By inspection of Eq.~\eqref{eq:ResProf:StrongDrivingScale},
we conclude that a larger driving amplitude $f_0$ generally favors satisfaction of Eq.~\eqref{eq:ResProf:StrongDriving:Condition}
since $\varphi_1(t), \varphi_2(t) \sim f_0^2$ according to~\eqref{eq:phi}.
Similary, a narrow perturbation profile in the sense of a faster decay of $\varV(E)$ with $E$ is expected to improve compliance with Eq.~\eqref{eq:ResProf:StrongDriving:Condition}
since it leads to smaller values of $\Sigma_0$,
which reduces the right-hand side of~\eqref{eq:ResProf:StrongDriving:Condition} more strongly than the left-hand side.
Broadly speaking,
we can thus conclude that
Eq.~\eqref{eq:ResProf:StrongDriving:GAux}
and its subsequent implications describe 
the cases of strong 
driving or narrow perturbation profiles $\varV(E)$,
bearing in mind that a more precise characterization must take into account the time dependence of Eq.~\eqref{eq:ResProf:StrongDriving:Condition} induced by the driving protocol in the setup of interest.

Substituting~\eqref{eq:ResProf:StrongDriving:GAux} into~\eqref{eq:ResProfFromG} yields
\begin{equation}
	\gamma(t, t') = \frac{2}{\pi r(t')^2} \int_{-r(t')}^{r(t')} \d E \; \e^{\I E t} \sqrt{r(t')^2 - E^2} \,.
\end{equation}
Evaluating the Fourier integral leads to 
\begin{equation}
\label{eq:ResProf:StrongDriving}
	\gamma(t, t') = \frac{ 2 J_1(r(t') t) }{ r(t') t }
\end{equation}
with $J_1(x)$ the Bessel function of the first kind of order 1.
For fixed $t'$, $\gamma(t, t')$ hence shows an oscillating decay to zero as a function of $t$
with a characteristic time scale proportional to $r(t')^{-1}$.
According to (\ref{eq:ResProf:StrongDrivingScale}), the
driving protocol $f(t)$
mediates a modulation of that time scale via 
$\varphi_1(t)$ and $\varphi_2(t)$ from
\eqref{eq:phi}.

\subsection{Fast driving}
\label{sec:ResProf:FastDriving}

The second approximation applies in the limit of sufficiently fast driving.
Here we start from the representation~\eqref{eq:ResProfEq} of $\gamma(t, t')$
and set $\varphi_2(t') \equiv 0$,
which corresponds to
truncating the Magnus expansion at first order \cite{dab24}
(as opposed to the second-order truncation that leads to~\eqref{eq:ResProfEq}).
In line with the common perception of
the Magnus expansion as a high-frequency approximation \cite{bla09},
the prediction resulting from such a lower-order truncation will work better for faster driving,
i.e., a smaller intrinsic time scale of $f(t)$;
see also Sec.~\ref{sec:ResProf:Num:Fast} for a more detailed discussion.

An approximate solution of
equation~\eqref{eq:ResProfEq}
with $\varphi_2(t') \equiv 0$ 
for rather general perturbation profiles can then be obtained 
along similar lines as in Ref.~\cite{dab21typical},
yielding
\begin{widetext}
\begin{equation}
\label{eq:ResProf:HighFreqDriving}
	\gamma(t, t') = \frac{ [r_1(t') - \hat r(t')] \e^{-r_{-1}(t') \lvert t \rvert} 
	- 2 \hat r(t') \e^{-r_0(t') \lvert t \rvert} + [r_{-1}(t') - \hat r(t')] \e^{-r_1(t') \lvert t \rvert} }{2 [r_0(t') - 2 \hat r(t')]}
\end{equation}
\end{widetext}
with
\begin{align}
	\hat r(t) &:= \pi \, \varV(0) \,\varphi_1(t) \, D_0 \,, \\
\label{eq:ResProf:HighFreqDriving:rn}
	r_n(t) &:= \frac{\Sigma_0}{\pi} \left( 1 + n \sqrt{1 - \frac{2 \pi \, \hat r(t)}{\Sigma_0} } \right) .
\end{align}
Note that $r_0(t')$ 
appearing in (\ref{eq:ResProf:HighFreqDriving}) is actually independent of $t'$
according to \eqref{eq:ResProf:HighFreqDriving:rn},
and that
this approximation depends solely on the two characteristic parameters 
$\varV(0) D_0$ and $\Sigma_0$ of the driving operator $V$.
Incidentally,
an even simpler approximation, $\gamma(t, t') = \e^{-\hat r(t') \lvert t \rvert}$, 
can be derived by the same methods as in Ref.~\cite{dab21typical} 
if, in addition, the driving amplitude is sufficiently small.

\begin{figure*}
\centering
\includegraphics[scale=1]{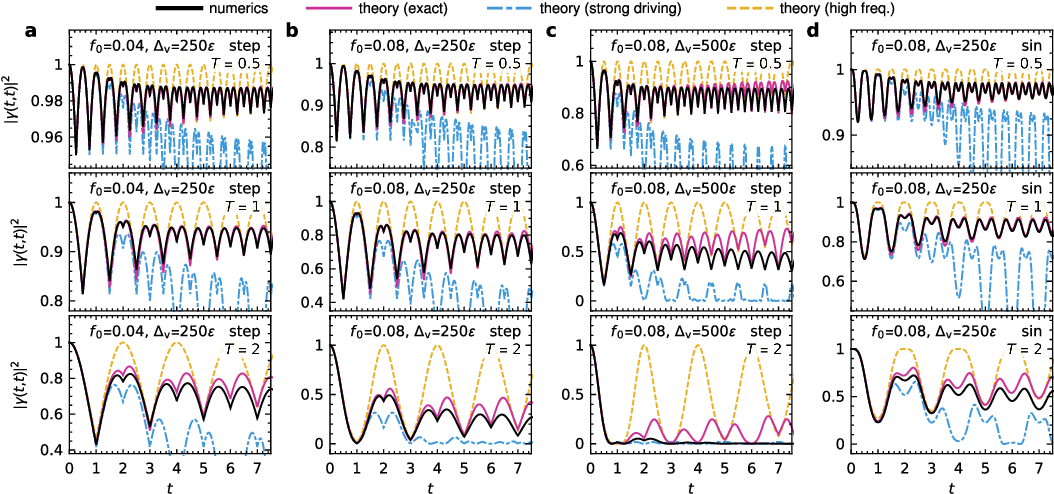} 
\caption{Squared response function $\lvert \gamma(t, t) \rvert^2$ 
versus $t$
for various perturbation band widths $\Delta_v$ and driving protocols $f(t)$
with amplitudes $f_0$ and periods $T$
as indicated in the top of each panel
(note the different $y$-axis scales;
the labels ``step'' and ``sin'' refer to (\ref{eq:DrivingProtocol:Step}) 
and (\ref{eq:DrivingProtocol:Sin}), respectively).
The black lines correspond to the 
numerically exact
solution of the Schr\"odinger equation for 
the eigenstate fidelity of a random matrix model 
with perturbation profile $\varV(E) = \varV(0) \, \e^{-\lvert E \rvert / \Delta_v}$
(see main text around~\eqref{eq:TypTimeEvoFidelity} for more details).
The pink lines (sometimes hardly distinguishable from the black lines)
are the corresponding 
theoretical prediction~\eqref{eq:TypTimeEvoFidelity}, obtained
by numerical integration of Eq.~\eqref{eq:ResProfEq}.
The yellow dashed lines correspond to the high-frequency approximation~\eqref{eq:ResProf:HighFreqDriving}.
The blue dash-dotted lines are the approximation~\eqref{eq:ResProf:StrongDriving} 
for strong driving
and
narrow perturbation profiles.
}
\label{fig:ResProf}
\end{figure*}

\subsection{Numerical verification}
\label{sec:ResProf:Num}

To verify and test the validity regimes of Eqs.~\eqref{eq:TypTimeEvo}--\eqref{eq:ResProfEq} and the just-derived approximations for $\gamma(t, t')$,
we consider a random-matrix model for the Hamiltonian $H(t)$ from~\eqref{eq:H}.
It is a direct implementation of the typicality framework underlying the derivation of the theory~\eqref{eq:TypTimeEvo}--\eqref{eq:ResProfEq},
such that we avoid any spurious influences of imperfect modeling.
Furthermore, we choose the system size sufficiently large so that finite-size effects are negligible on the relevant scales.

The unperturbed Hamiltonian $H_0 = \sum_\mu E_\mu \ket{\mu} \bra{\mu}$
is defined via its energy levels $E_\mu = \mu \varepsilon$
with constant level spacing $\varepsilon = D_0^{-1} = 2^{-9} = 1/512$,
such that the (coarse-grained) density of states 
(see above Eq.~\eqref{eq:Auth})
is homogeneous
across all of the $M = 2^{14} = 16\,384$ levels.
The driving operator $V$ is a random Hermitian matrix with complex Gaussian-distributed 
entries $V_{\mu\nu}$ of vanishing mean and variance
\begin{equation}
\label{eq:VarV:Exp}
	\varV(E) = \varV(0) \, \e^{-\lvert E \rvert / \Delta_v} \,.
\end{equation}
We focus on periodic driving protocols and consider a step shape,
\begin{equation}
\label{eq:DrivingProtocol:Step}
	f(t) = f_0 \, \operatorname{sgn}[ \sin(2\pi t / T) ]\,,
\end{equation}
where $\operatorname{sgn}(x)$ denotes the sign function,
and a sinusoidal shape
\begin{equation}
\label{eq:DrivingProtocol:Sin}
	f(t) = f_0 \, \sin(2\pi t / T) \,.
\end{equation}
The initial state $\rho(0) = \ketN{\alpha}\braN{\alpha}$ is an eigenstate of $H_0$ from 
the middle of the spectrum ($\alpha = M/2 = 8192$).
Choosing $A = \rho(0)$, the time-dependent expectation values 
$\< A \>_{\!\rho(t)} = \lvert \braN{\alpha} \mc U(t) \ketN{\alpha} \rvert^2$ monitor the fidelity or survival probability of the initial state.
In particular, we thus have $\< A \>_{\!\rho_0(t)} = 1$ and $\Auth \simeq 0$.
Hence,
our theoretical prediction~\eqref{eq:TypTimeEvo} boils down to
\begin{equation}
\label{eq:TypTimeEvoFidelity}
\< A \>_{\!\rho(t)} = \lvert \gamma(t, t) \rvert^2
\,,
\end{equation}
meaning that we can directly compare the 
numerically time-evolved expectation values 
$\< A \>_{\!\rho(t)}$ to the 
solutions of Eq.~\eqref{eq:ResProfEq}.

These time-evolved expectation values $\< A \>_{\!\rho(t)}$ thus represent the ground truth.
They are shown as black lines for various choices of the driving amplitude $f_0$, the driving period $T$, 
and the perturbation band width $\Delta_v$ in Fig.~\ref{fig:ResProf}.
Note 
that the response behavior is clearly far from being linear
in the driving amplitude $f_0$
and thus is outside the realm of what could 
be captured by linear response theory.

\begin{figure*}
\includegraphics[scale=1]{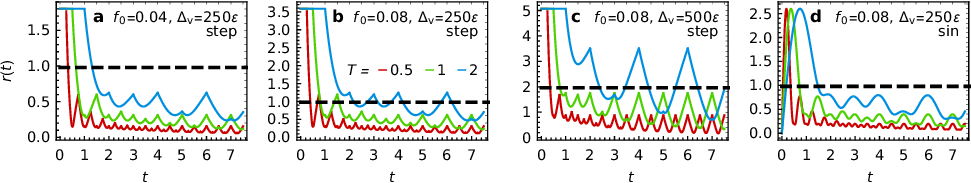}
\caption{Characteristic energy/inverse-time scale $r(t)$ [cf.\ Eq.~\eqref{eq:ResProf:StrongDrivingScale}] of the analytical approximation~\eqref{eq:ResProf:StrongDriving} of $\gamma(t, t')$ for large driving amplitudes 
and
narrow perturbation profiles.
The perturbation profile $\varV(E)$ and driving protocol $f(t)$ in panels (a)--(d) are in correspondence to the respective panels of Fig.~\ref{fig:ResProf},
as indicated in the top-right corners.
Colors correspond to the different driving periods, see the legend in (b).
The dashed black line marks the value of $\Sigma_0$, i.e., the right-hand side of the condition~\eqref{eq:ResProf:StrongDriving:Condition}.}
\label{fig:ResProf:StrongDrivingScale}
\end{figure*}

\subsubsection{Exact solutions of Eq.~\eqref{eq:ResProfEq}}

We first compare these $\< A \>_{\!\rho(t)}$ to the numerically exact solutions for $\gamma(t, t')$,
which we obtain by integrating Eq.~\eqref{eq:ResProfEq} numerically.
The resulting $\lvert \gamma(t, t) \rvert^2$ are
shown as pink lines in Fig.~\ref{fig:ResProf}.
We recall that the relations~\eqref{eq:TypTimeEvo} and~\eqref{eq:ResProfEq} were derived in Ref.~\cite{dab24} by adopting a second-order Magnus expansion,
which naturally limits the applicability to an initial transient time window.
This window is expected to grow as the characteristic time scale of the driving (i.e., the period $T$ in the present example) becomes smaller.
In Fig.~\ref{fig:ResProf},
we observe that the simulated dynamics (black lines) and the theoretical prediction (pink lines) indeed agree excellently in the expected regime,
i.e., for short times and/or small driving periods.
We emphasize that there is no fitting involved because all parameters entering the theoretical prediction~\eqref{eq:TypTimeEvo}--\eqref{eq:ResProfEq} are known by construction.

As mentioned above, the typicality framework in the derivation of our main result
is essentially exact for the present random-matrix model.
Hence
the remaining deviations in Fig.~\ref{fig:ResProf} -- 
seen particularly for larger driving periods (smaller frequencies) and late times 
-- must 
be attributed to the employed truncation of the Magnus 
expansion at second order.

By comparing Figs.~\ref{fig:ResProf}b and d,
we observe
that the shape of the driving profile (here: step vs.\ sinusoidal) appears to be 
of minor relevance with respect to the prediction's accuracy 
(deviations become noticeable at similar times when all other parameters are fixed).
Similarly, a comparison of Figs.~\ref{fig:ResProf}a and b suggests that the influence 
of the driving amplitude $f_0$ on the accuracy is relatively small.
(We remark
that the crossover between the ``weak'' and ``strong'' perturbation regimes 
can be shown \cite{dab20modification} to occur
at $f_0 \simeq \sqrt{2\varepsilon\Delta_v / \pi^2 \varV(0)}$.
For the parameters in Fig.~\ref{fig:ResProf},
this value lies in the range $0.01 \ldots 0.02$,
meaning that, in this sense, all 
depicted curves already correspond
to relatively ``strong'' drivings,
though the effective strength is ultimately time dependent,
see also the discussion below Eq.~\eqref{eq:ResProf:StrongDriving:Condition}.)

On the other hand,
the width $\Delta_v$ of the perturbation profile $\varV(E)$ from~\eqref{eq:VarV} and~\eqref{eq:VarV:Exp}
significantly affects the quality of the approximation,
as indicated by comparing Figs.~\ref{fig:ResProf}b and c.
Since $\Delta_v$ roughly quantifies the energy range of the perturbation (i.e., the 
scale across which eigenstates of the unperturbed system $H_0$ are coupled directly by the driving operator $V$),
reducing
$\Delta_v$ suppresses transitions between levels that are far apart in the unperturbed spectrum.
As a consequence, broadening of the state $\rho(t)$ and thus heating occur more slowly,
facilitating a longer reliability of the truncated Magnus expansion.

In summary,
both small driving periods $T$ and narrow perturbation profiles 
$\varV(E)$ have a favorable effect on the applicability of the theory.
By contrast, the magnitude of the driving amplitude $f_0$ is less important,
particularly meaning that quite large values
can still be acceptable.
Qualitatively,
the effective
magnitude of $\lvert \gamma(t, t) \rvert^2$ 
(e.g., time-averaged over one driving period $T$)
always decreases with increasing
$T$.
For asymptotically small driving periods, 
in turn, the theory predicts $\lvert \gamma(t, t) \rvert^2 \simeq 1$
for all $t$, 
and the numerics in Fig.~\ref{fig:ResProf} (note the $y$-axis scales)
confirms that the system is then indeed
(as one might have expected) unable to follow the 
rapidly oscillating driving and instead essentially reproduces the reference 
dynamics $\< A \>_{\!\rho_0(t)}=1$ induced by the time-averaged 
Hamiltonian $H_0$.
Moreover,
an obvious and repeatedly emphasized observation is that faster driving (smaller $T$) 
leads to longer agreement with the true dynamics (black lines) for all theory curves in Fig.~\ref{fig:ResProf},
a direct consequence of the truncated Magnus expansion.

Finally,
we
compare the analytical approximations from Secs.~\ref{sec:ResProf:StrongDriving} and~\ref{sec:ResProf:FastDriving}
to the numerically exact solution of Eq.~\eqref{eq:ResProfEq} (pink lines) and to the actual response characteristics of the random-matrix model (black lines).

\subsubsection{Fast driving}
\label{sec:ResProf:Num:Fast}

We first inspect the high-frequency approximation~\eqref{eq:ResProf:HighFreqDriving} from Sec.~\ref{sec:ResProf:FastDriving},
shown as dashed yellow lines in Fig.~\ref{fig:ResProf}.
As mentioned above, it is tantamount to a first-order Magnus expansion,
setting $\varphi_2(t') \equiv 0$ in Eq.~\eqref{eq:ResProfEq}.

Our first conclusion is
that the second-order Magnus expansion (solid pink lines)
provides a significant improvement
over the first-order approximation.
For the periodic driving protocols considered in Fig.~\ref{fig:ResProf},
this manifests itself particularly at integer multiples of the driving period,
where the high-frequency, first-order approximation always returns to 
its initial value $1$.
This follows from the fact that $\varphi_1(nT) = 0$ for $n \in \NN$ and the periodic, unbiased driving protocols chosen in Fig.~\ref{fig:ResProf},
such that $\gamma(t, nT) = 1$ in this approximation.

On the other hand,
around half-integer multiples of the driving period [$t = (n-\frac{1}{2}) T$],
the high-frequency approximation (dashed yellow lines) agrees very well with the exact solutions of Eq.~\eqref{eq:ResProfEq} (pink lines).
This can be understood by observing that $\varphi_2([n-\frac{1}{2}]T) = 0$ for $n \in \NN$ and the chosen driving protocols,
such that the first- and second order approximations for $\gamma(t, [n-\frac{1}{2}]T)$ agree.
Thanks to this agreement at half-integer multiples of $T$,
the analytical first-order approximation remains valuable to estimate the overall strength of the response on similar time scales as the full second-order solutions of Eq.~\eqref{eq:ResProfEq},
even though it misses quantitative details around integer multiples of $T$.

The high-frequency character of the approximation~\eqref{eq:ResProf:HighFreqDriving}
is apparent by comparing the time up to which the approximation stays below a preset error threshold
to the intrinsic time scale of the response (e.g., the period $T$).
Focusing on Fig.~\ref{fig:ResProf}a, for example,
the relative error (deviation divided by initial amplitude) after two periods in the top panel is about the same as the relative error after one period in the middle panel.
Likewise, the relative error after two periods in the middle panel is smaller than the relative error after one period in the bottom panel.
Similar conclusions can be drawn from Figs.~\ref{fig:ResProf}b--d as well.

\subsubsection{Strong and short-ranged-in-energy driving}
\label{sec:ResProf:Num:Strong}

The approximation~\eqref{eq:ResProf:StrongDriving} for strong driving 
and
quickly 
decaying perturbation profiles from Sec.~\ref{sec:ResProf:StrongDriving} is shown as dash-dotted blue lines in Fig.~\ref{fig:ResProf}.
It yields good agreement for small times,
but shows larger deviations and, in particular, a distinct asymptotic behavior for later times.
To understand this,
we inspect the scale $r(t)$ from Eq.~\eqref{eq:ResProf:StrongDrivingScale},
which controls the quality of the approximation according to~\eqref{eq:ResProf:StrongDriving:Condition}.
It is shown in Fig.~\ref{fig:ResProf:StrongDrivingScale} for the same parameters as in Fig.~\ref{fig:ResProf}.
As discussed below Eq.~\eqref{eq:ResProf:StrongDriving:Condition},
the approximation~\eqref{eq:ResProf:StrongDriving} is only expected to work if $r(t)$ significantly exceeds the value of $\Sigma_0$ (black dashed lines in Fig.~\ref{fig:ResProf:StrongDrivingScale}),
cf.\ Eq.~\eqref{eq:ResProf:StrongDriving:Condition}.

We notice that,
for the present periodic driving protocols,
$r(t)$ is significantly larger during the first period compared to the fluctuations observed at later times.
Consequently, the approximation here comes with an additional preference for short times on top of the limitations resulting from truncating the Magnus expansion.

Moreover, a similar trend as observed above when comparing the numerically exact solutions of Eq.~\eqref{eq:ResProfEq} (pink lines) to the ground truth (black lines) becomes apparent here:
For the present parameter values,
the quality of the approximation is overall more sensitive to the width $\Delta_v$ of the perturbation profile than to the amplitude $f_0$ of the driving.

Interestingly,
the approximation~\eqref{eq:ResProf:StrongDriving} is somewhat complementary to the first-order approximation from~\eqref{eq:ResProf:HighFreqDriving}
in the sense that when the latter overestimates $\lvert \gamma(t, t) \rvert^2$, the former usually underestimates it.

\begin{figure*}
\includegraphics[scale=1]{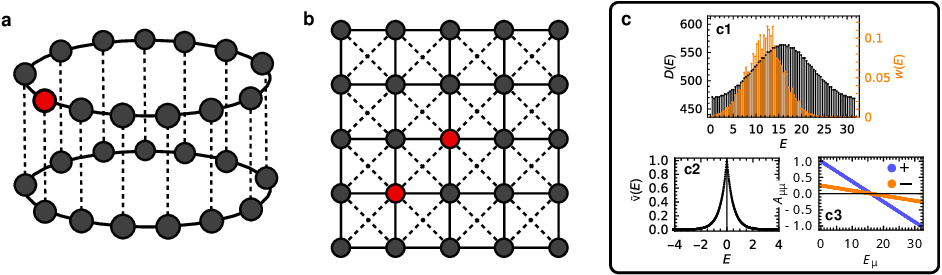}
\caption{Sketches illustrating some key features of the three example systems from Secs.~\ref{sec:FiniteQuench}--\ref{sec:DoublePretherm}.
(a)~Heisenberg spin rings from Sec.~\ref{sec:FiniteQuench}. The unperturbed system~\eqref{eq:Spin12x2p:H0} consists of two isolated rings 
(solid links), and the
driving operator~\eqref{eq:Spin12x2p:V} connects the rings site-wise (dashed links).
The observable measures a single-site magnetization (red).
(b) Two-dimensional spin lattice with open boundary conditions from Sec.~\ref{sec:PseudrandomDriving}.
The unperturbed system~\eqref{eq:Spin5x5:H0} exhibits couplings between
nearest neighbors (solid links), and the driving operator~\eqref{eq:Spin5x5:V}
additionally connects next-nearest neighbors (dashed links).
The observable~\eqref{eq:mc} measures the spin correlation between
the two red sites.
(c) Random matrix model as introduced in Sec.~\ref{sec:DoublePretherm}. Subpanels show (c1) the (coarse-grained) density of states $D(E)$ (see above Eq.~\eqref{eq:Auth}) of the reference Hamiltonian $H_0$ (in black) and the (coarse-grained) energy distribution $w(E) := \sum_\mu \rho_{\mu\mu}(0) \, \delta(E - E_\mu)$ of the initial state (in orange); (c2) the perturbation profile~\eqref{eq:VarV}; (c3) the observable's diagonal matrix elements in the `$+$' and `$-$' sectors (cf.\ Eq.~\eqref{eq:S:RMMPretherm:A}).}
\label{fig:ExampleModels}
\end{figure*}

\begin{figure*}
\includegraphics[scale=0.85]{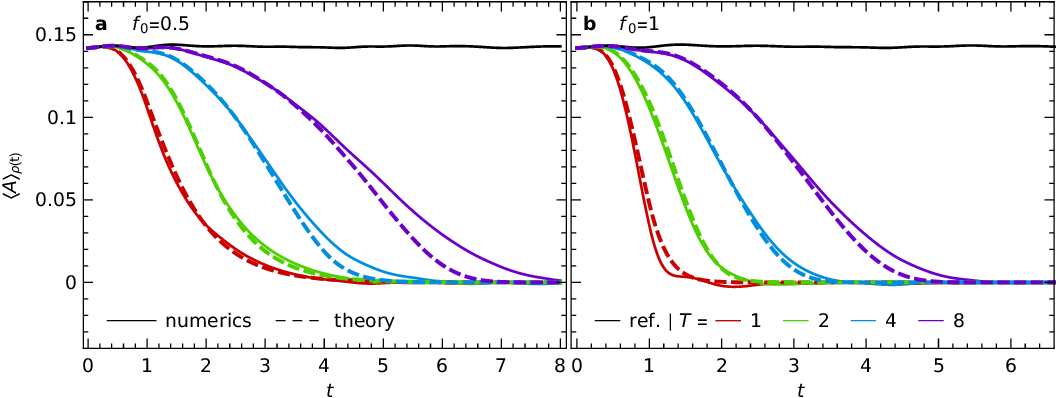}
\caption{Time-dependent expectation values $\< A \>_{\!\rho(t)}$ of the single-site 
magnetization $A = \sigma^z_{1,1}$
in the spin-ring system~\eqref{eq:Spin12x2p:H0} with $L = 14$ subject to the linear-ramp protocol (finite-time quench) from~\eqref{eq:DrivingProtocol:LinearRamp},
which couples the two initially isolated rings via~\eqref{eq:Spin12x2p:V},
cf.\ the sketch in Fig.~\ref{fig:ExampleModels}a.
The initial state~\eqref{eq:InitStateQuench} emulates a thermal equilibrium state with magnetizations $\< S^z_1 \>_{\!\rho(0)} = -\< S^z_2 \>_{\!\rho(0)} = 2$.
In each panel, data for quench times
$T = 1, 2, 4, 8$
are shown 
along with the unperturbed reference dynamics $\< A \>_{\!\rho_0(t)}$ 
as indicated in the legend of (b);
the remnant oscillations of the unperturbed dynamics (black curves) are finite-size effects \cite{fig:Spin12x2pQuench:FiniteSize}.
Solid lines:
numerical results.
Dashed lines:
theoretical prediction~\eqref{eq:TypTimeEvo}, 
adopting the numerically obtained undriven behavior $\< A \>_{\!\rho_0(t)}$ (black curves),
squared response function $\lvert \gamma(t, t) \rvert^2$ (by numerical integration of~\eqref{eq:ResProfEq}), 
and the 
thermal
value $\Auth = 0$ (see below Eq.~(\ref{eq:InitStateQuench})).
}
\label{fig:Spin12x2pQuench}
\end{figure*}

\section{Finite-time quenches}
\label{sec:FiniteQuench}

The driving protocol $f(t)$ in~\eqref{eq:H}
is not restricted to periodic functions,
but can take largely arbitrary shapes
as long as it remains reasonable to connect the driven Hamiltonian $H(t)$ to the unperturbed $H_0$ at any fixed time $t$.
Another common scenario of time-dependent forcing is a so-called \emph{quench},
where the system is prepared in a stationary state of one Hamiltonian $H_0$ 
and subsequently time-evolved with another Hamiltonian $H'$.
The necessary parameter change to switch from $H_0$ to $H'$ could involve changing an external field or bringing two previously isolated systems into contact, for example.
This change is often modeled to occur instantaneously,
but will take a finite amount of time in any real system.
Our theory~\eqref{eq:TypTimeEvo}--\eqref{eq:ResProfEq} allows us to predict the dynamics during and after such finite-time quenches,
provided that they are sufficiently weak and still happen sufficiently fast.

To get a qualitative understanding of the predictions in a quench-like scenario,
we consider a linear-ramp protocol
\begin{equation}
\label{eq:DrivingProtocol:LinearRamp}
	f(t) = f_0 \, \Theta(t) \left[ \frac{t}{T} \Theta(T - t) + \Theta(t - T) \right] ,
\end{equation}
where $\Theta(t)$ is the Heaviside step function.
In view of (\ref{eq:H}),
such a choice of
$f(t)$ thus
mediates
a linear interpolation between the pre-quench Hamiltonian 
$H_0$ (reference system) 
and the post-quench Hamiltonian 
\begin{equation}
H' = H_0 + f_0 V
\label{neu2}
\end{equation}
in time $T$.
Other shape functions for the quench protocol
will behave qualitatively similarly
as far as the following discussion is concerned.

Focusing on the post-quench dynamics first,
we consider late times $t \gg T$
and find 
from~\eqref{eq:F}, \eqref{eq:phi}, and \eqref{eq:DrivingProtocol:LinearRamp} 
that $\varphi_1(t) \sim f_0^2$ and $\varphi_2(t) \sim f_0^2 T^2 / 16$.
Together with Eq.~\eqref{eq:ResProfEq},
we thus conclude that,
for short quench times ($T \to 0$), 
the late time behavior is similar to the one observed after an instantaneous quench,
and we recover as a special case
the theory for time-independent perturbations from Refs.~\cite{dab20relax, dab21typical}.
The slower the quench,
the more important does the finite-time correction due to the term proportional to $\varphi_2(t')$ in~\eqref{eq:ResProfEq} become,
and eventually also higher orders of the Magnus expansion will have to be taken into account.

Second, we consider the limit $t \ll T$ and thus inspect the dynamics at the initial stage during the quench.
Here we find 
from~\eqref{eq:F}, \eqref{eq:phi}, and \eqref{eq:DrivingProtocol:LinearRamp} that
$\varphi_1(t) \sim (f_0 t / 2T)^2$ and $\varphi_2(t) \sim (f_0 t^2 / 12 T)^2$.
From~\eqref{eq:ResProfEq}, we conclude that the rate at which $\gamma(t, t')$ changes with $t$ becomes smaller with increasing $T$.
As one might have expected intuitively, 
the dynamics is thus effectively slowed down compared to an instantaneous quench.

To illustrate these findings and verify them in a physical example,
we consider a linear quench~\eqref{eq:DrivingProtocol:LinearRamp} in a setup with two Heisenberg 
spin rings of $L = 14$ spins each, sketched in Fig.~\ref{fig:ExampleModels}a.
The rings are isolated from each other (and the outside world)
in the unperturbed (pre-quench) system,
\begin{equation}
\label{eq:Spin12x2p:H0}
	H_0 = H^{(1)} + H^{(2)} \,, \quad
	H^{(s)} := \sum_{i=1}^L \bm\sigma_{s,i} \cdot \bm\sigma_{s,i+1} \,,
\end{equation}
and interact sitewise in the driven system 
(during and after the quench)
according to
\begin{equation}
\label{eq:Spin12x2p:V}
	V = \sum_{i=1}^L \bm\sigma_{1,i} \cdot \bm\sigma_{2,i} \,,
\end{equation}
see dashed lines in Fig.~\ref{fig:ExampleModels}a.
The symbol $\bm\sigma_{s,i} := (\sigma^x_{s,i}, \sigma^y_{s,i}, \sigma^z_{s,i})$ 
denotes a vector of Pauli matrices acting on the $i$th site of the $s$th chain.

In the pre-quench system $H_0$ from~\eqref{eq:Spin12x2p:H0}, the magnetizations 
$S^\alpha_{s} := \sum_{i=1}^L \sigma^\alpha_{s,i}$ 
($\alpha = x,y,z$) of the two chains ($s = 1, 2$) 
are thus conserved individually, while in the post-quench 
system $H'$ from
(\ref{neu2})
only the total magnetization 
$S^\alpha := S^\alpha_1 + S^\alpha_2$ is still conserved.
In the following, we focus on the dynamics in the 
subsector with vanishing $S^z$.

As our observable, we consider the magnetization $A = \sigma^z_{1,1}$ of the first chain's 
first spin in the $z$ direction
(red dot in Fig.~\ref{fig:ExampleModels}a).
The initial state
is of the form
\begin{equation}
\label{eq:InitStateQuench}
	\rho(0) = \ket\psi \! \bra\psi \;\text{ with }\; \ket\psi \propto \e^{-(H_0 - E)^2 / 4 \Delta E^2} \, 
	{\ket{\phi}},
\end{equation}
where
$\ket{\phi}$ is a Haar-random state in the subsector
with magnetizations $S^z_1 = 2$ and $S^z_2 = -2$.
The subsequent Gaussian filter in~\eqref{eq:InitStateQuench} with
$E = -14$ and $\Delta E = 4$ ensures a macroscopically well-defined 
energy as required in the beginning of Sec.~\ref{sec:Theory:Setup}.
This initial state~\eqref{eq:InitStateQuench} thus emulates a
thermal equilibrium state of the two isolated spin rings,
but results in a nonequilibrium state when bringing the rings into contact via the quench.
Hence the expectation value $\< A \>_{\!\rho_0(t)}$ 
of the unperturbed reference system is constant, 
$\< A \>_{\!\rho_0(t)} = \Autav = 2/L$ \cite{fig:Spin12x2pQuench:FiniteSize}.
As a result of the quench, we expect the dynamics $\< A \>_{\!\rho(t)}$ 
of the joint system to approach 
the post-quench thermal equilibrium value of the interacting chains, $\Auth = 0$ (since the total magnetization of the joint system of both chains is $S^z = 0$).

For the theoretical prediction according to Eqs.~\eqref{eq:TypTimeEvo} and~\eqref{eq:ResProfEq},
we assume an exponential shape~\eqref{eq:VarV:Exp} for the perturbation profile and estimate the parameters $\varV(0) D_0 = 0.98$ 
and $\Delta_v = 4.2$ from the corresponding instantaneous-quench scenario using 
the theory of Ref.~\cite{dab21typical} (see also Ref.~\cite{dab24}).

\begin{figure*}[t]
\includegraphics[scale=1]{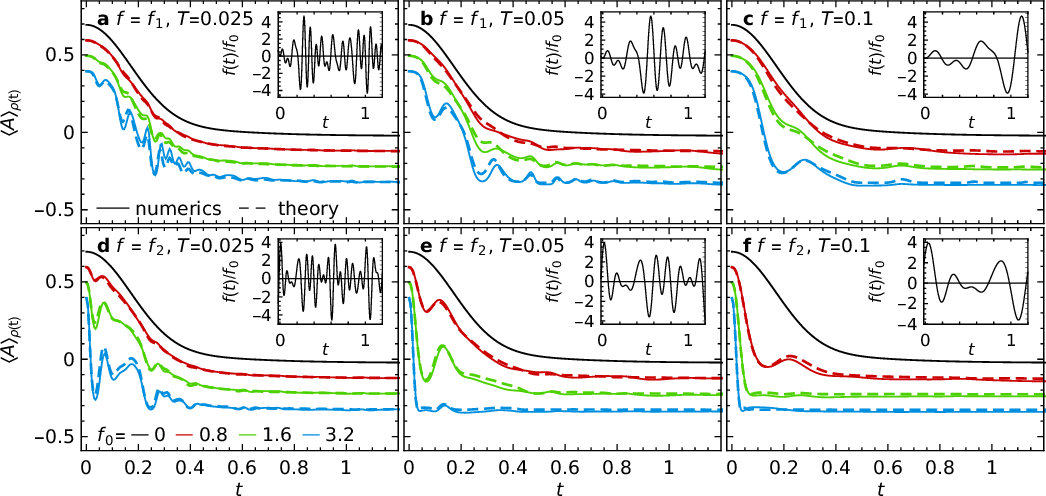}
\caption{Time-dependent expectation values $\< A \>_{\!\rho(t)}$ of the magnetization 
correlation $A$ from~\eqref{eq:mc} for the two-dimensional spin model 
from~\eqref{eq:Spin5x5:H0}--\eqref{eq:Spin5x5:V}
(see also Fig.~\ref{fig:ExampleModels}b),
driven pseudorandomly according to~\eqref{eq:Driving:Pseudorand1}--\eqref{eq:Driving:Pseudorand2} (see insets) for various values of $T$ (see top left corner of each panel) and $f_0$ (see legend in (d)).
The initial state is of the form~\eqref{eq:InitState} with $Q = \pi^+_{2,2} \pi^+_{3,3}$, $E = -12$, and $\Delta E = 4$.
The $y$ values for $f_0 = 0.8, 1.6, 3.2$ are shifted in steps of $0.1$ for the sake of clarity.
Solid lines show 
numerical results
obtained via Suzuki-Trotter propagation.
Dashed lines correspond to the theoretical prediction~\eqref{eq:TypTimeEvo}
utilizing the numerically obtained undriven behavior $\< A \>_{\!\rho_0(t)}$
(black curves),
the function $\lvert \gamma(t, t) \rvert^2$ calculated by numerical integration of~\eqref{eq:ResProfEq},
and the 
thermal equilibrium (microcanonical) 
expectation value $\Auth = -0.026$ for the initially occupied energy window.
}
\label{fig:Pseudorand}
\end{figure*}

As shown in Fig.~\ref{fig:Spin12x2pQuench}, 
our theoretical prediction~\eqref{eq:TypTimeEvo} explains the
numerical results very well and without any 
adjustable fitting parameter.
As anticipated below Eq.~\eqref{eq:ResProfEq} and in the beginning of this section,
the agreement between theory and numerics is best if $t$ and/or $T$ are small,
which is a consequence of the truncated Magnus expansion in the underlying 
derivation.
Quantitatively, the supported $t$ and $T$ values are still
remarkably large,
and the adopted quench amplitudes $f_0$ are on the same order as the intrinsic interactions in the pre-quench system.

Overall, the example in Fig.~\ref{fig:Spin12x2pQuench} confirms and quantitatively
illustrates the basic features of (sufficiently weak) finite-time quenches as predicted by the 
theory~\eqref{eq:TypTimeEvo}--\eqref{eq:ResProfEq}:
On the one hand, the relaxation is slowed down upon increasing the quench time $T$,
which implies, in particular, that the dynamics $\< A \>_{\!\rho(t)}$ of the quenched system
remains closer to the unperturbed equilibrium value $\< A \>_{\!\rho_0(t)} = \Autav$
at short times.
On the other hand,
the quenched dynamics $\< A \>_{\!\rho(t)}$ eventually approaches the same (thermal) 
value at late times for all (sufficiently small) $f_0$ and $T$.

We finally note that the observed relaxation dynamics 
in our present example, where two previously isolated systems are brought into
contact by the perturbation, is once again beyond of what could possibly be 
understood in terms of some linear response theory.

\section{Pseudorandom driving}
\label{sec:PseudrandomDriving}

To illustrate the diversity of setups that the theory can describe,
our next application involves two ``pseudorandom'' driving protocols
$f(t)$ in (\ref{eq:H}), namely
\begin{equation}
\label{eq:Driving:Pseudorand1}
	f(t)=f_1(t) := f_0 \sum_{k=1}^6 (-1)^{k+1} \cos\!\left( \frac{\sqrt{k} \, t}{T} \right) \,.
\end{equation}
and
\begin{equation}
\label{eq:Driving:Pseudorand2}
	f(t)=f_2(t) := f_0 \sum_{k=1}^3 \left[ \sin\!\left( \frac{\sqrt{2k-1} \, t}{T} \right) + \cos\!\left( \frac{\sqrt{2k} \, t}{T} \right) \right] .
\end{equation}
Plots of these protocols for different values of the time scale $T$ can be found in the insets of Fig.~\ref{fig:Pseudorand}.

For the undriven reference system, we choose a spin-$\frac{1}{2}$ model on a two-dimensional square lattice with 
nearest-neighbor Heisenberg interactions,
\begin{equation}
\label{eq:Spin5x5:H0}
	H_0 = \sum_{i=1}^L \sum_{j = 1}^{L-1} \left( \bm\sigma_{i,j} \cdot \bm\sigma_{i,j+1} + \bm\sigma_{j,i} \cdot \bm\sigma_{j+1, i} \right) \,,
\end{equation}
with $L = 5$;
see also the sketch in Fig.~\ref{fig:ExampleModels}b.
The driving operator $V$ couples next-nearest neighbors via spin-flip terms in the $z$ direction,
\begin{equation}
\label{eq:Spin5x5:V}
	V = \sum_{i,j=1}^{L-1} \sum_{\alpha=x,y} (\sigma^\alpha_{i,j} \sigma^\alpha_{i+1,j+1} + \sigma^\alpha_{i,j+1} \sigma^\alpha_{i+1,j}) \,,
\end{equation}
see dashed lines in Fig.~\ref{fig:ExampleModels}b.
The system is prepared in a state of the form
\begin{equation}
\label{eq:InitState}
	\rho(0) = \ket\psi \! \bra\psi \;\text{ with }\; \ket\psi \propto \e^{-(H_0 - E)^2 / 4 \Delta E^2} \, Q \, \ket\phi,
\end{equation}
where $\ket\phi$ is a Haar-random state in the subsector with eigenvalue $-1$ of the 
(conserved) $z$ magnetization $S^z := \sum_{i,j} \sigma^z_{i,j}$.
Furthermore,
the projector $Q = \pi^+_{2,2} \pi^+_{3,3}$
with $\pi_{i,j}^+ := (1 + \sigma^z_{i,j})/2$
deflects two spins near the center of the lattice in the positive $z$ direction
(see red dots in Fig.~\ref{fig:ExampleModels}b),
and the Gaussian filter with
$E = -12$ and $\Delta E = 4$ ensures a macroscopically well-defined energy as 
before (see below Eq. (\ref{eq:InitStateQuench})).
As our observable, we choose the magnetization correlation between the two initially deflected sites,
\begin{equation}
\label{eq:mc}
	A := \sigma_{2,2}^z \sigma_{3,3}^z \,,
\end{equation}
indicated by the red dots in Fig.~\ref{fig:ExampleModels}b.

The colored solid lines in Fig.~\ref{fig:Pseudorand} show the numerically obtained time-evolved expectation values for the two driving protocols $f_1(t)$ and $f_2(t)$ from Eqs.~\eqref{eq:Driving:Pseudorand1} and~\eqref{eq:Driving:Pseudorand2}, respectively, and various combinations of the parameters $f_0$ and $T$.
We observe a large variety of response characteristics in the initial dynamics, as long as the undriven system (black lines) is far away from equilibrium.

For a comparison with the theory~\eqref{eq:TypTimeEvo}--\eqref{eq:ResProfEq},
we assume again the exponential form~\eqref{eq:VarV:Exp} for the perturbation characteristic $\varV(E)$ from~\eqref{eq:VarV},
which has been verified empirically for a similar
but
 smaller system with $L = 4$ in~\cite{dab20modification}.
The
parameters $\varV(0) D_0 \simeq 3.6$ and $\Delta_v \simeq 3.0$ were
estimated independently from the relaxation of the system under a constant perturbation \cite{dab21typical}.
For given $f(t)$, we can then solve Eq.~\eqref{eq:ResProfEq} to find $\lvert \gamma(t, t) \rvert^2$,
and substitute it into~\eqref{eq:TypTimeEvo} together with the numerically computed undriven behavior $\< A \>_{\!\rho_0(t)}$ (black lines) and thermal expectation value $\Auth = -0.026$.

Remarkably, the resulting theoretical predictions, overlaid as dashed lines in Fig.~\ref{fig:Pseudorand}, reproduce the 
diverse features 
of the numerically exact results very well.
Both the vibrant oscillations for small values of $T$ and the smoother relaxation for larger values of $T$ build on the same (quasi-monotonous) undriven dynamics and are encoded very accurately in the solutions of Eq.~\eqref{eq:ResProfEq},
which uses the driving protocol $f(t)$ and (the Fourier transform of) the coarse-grained energy profile~\eqref{eq:VarV} of the driving operator as its only input parameters.

\section{Double prethermalization}
\label{sec:DoublePretherm}

\begin{figure}
\includegraphics[scale=1]{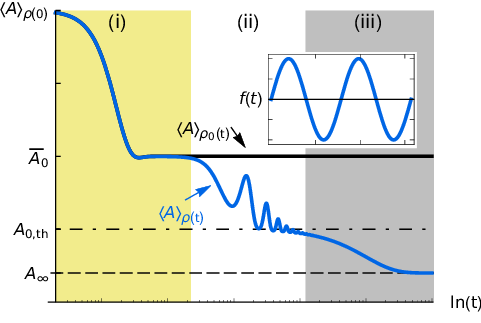}
\caption{Qualitative sketch of the time-dependent expectation values 
(note the logarithmic time scale) for
the double-prethermalization effect predicted by the theory~\eqref{eq:TypTimeEvo}
for systems subject to a symmetry-breaking periodic driving
(see inset).
The driven dynamics (blue curve) undergoes three stages:
(i) Relaxation to the long-term value $\Autav$ of the (nonthermalizing) undriven system (black 
line);
(ii) 
oscillations between $\Autav$ and
the thermal value $\Auth$ of the unperturbed system;
(iii) heating-induced
approach of the infinite-temperature value $\Ainf$.}
\label{fig:DoublePrethermSketch}
\end{figure}

\begin{figure*}
\includegraphics[scale=1]{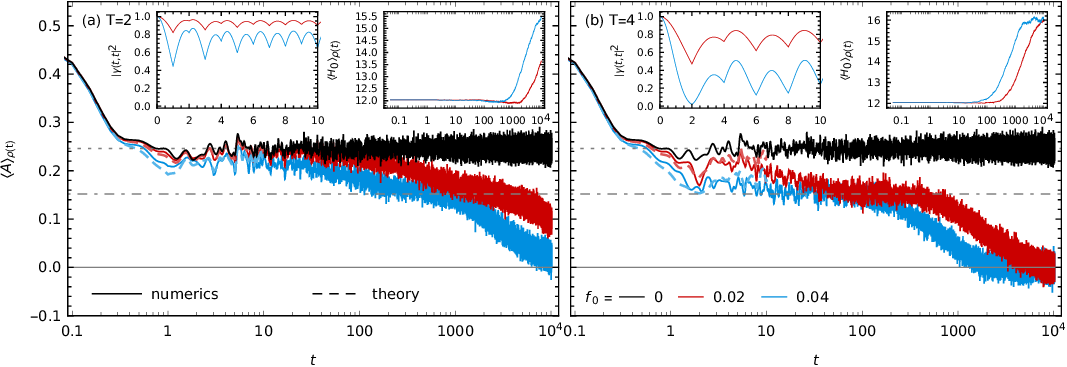}
\caption{Time-dependent expectation values $\< A \>_{\!\rho(t)}$ of the observable $A$ for the 
random-matrix model from Sec.~\ref{sec:DoublePretherm} and Fig.~\ref{fig:ExampleModels}c.
The driving protocol is of the periodic
step form~\eqref{eq:DrivingProtocol:Step} with (a)~$T = 2$ and (b)~$T = 4$, and 
amplitudes $f_0 = 0, 0.02, 0.04$ as indicated in the legend 
of~(b).
Solid lines show 
numerical results
obtained via exact diagonalization.
Dashed bright lines (shown for $t \leq 10$) correspond to the theoretical prediction~\eqref{eq:TypTimeEvo}
utilizing the numerically obtained undriven behavior $\< A \>_{\!\rho_0(t)}$
(black curves),
the function $\lvert \gamma(t, t) \rvert^2$ calculated by numerical integration of~\eqref{eq:ResProfEq},
and the 
thermal equilibrium (microcanonical) 
expectation value $\Auth = 0.15$ for the initially occupied energy window (dash-dotted horizontal line).
At long times, the unperturbed system approaches $\Autav = 0.25$ (dotted horizontal line), while the driven dynamics $\< A \>_{\!\rho(t)}$ approaches the infinite-temperature value $\Ainf = 0$ (solid horizontal line).
Insets: The function $\lvert \gamma(t,t) \rvert^2$, characterizing the response 
to the driving according to \eqref{eq:TypTimeEvo},
and the 
expectation
value $\< H_0 \>_{\!\rho(t)}$ of the unperturbed reference Hamiltonian,
which quantifies the energy exchange (heating)
caused by the periodic driving.
}
\label{fig:S:RMMPrepretherm}
\end{figure*}

One remarkable phenomenon predicted by the theory~\eqref{eq:TypTimeEvo}--\eqref{eq:ResProfEq}
is the ``stalled response'' effect which we extensively discussed in Ref.~\cite{dab24}:
A thermalizing many-body system reacts much more strongly to time-periodic perturbations if the accompanying undriven/time-averaged system is far from thermal equilibrium.
As the system approaches a thermal state (or if it finds itself close to it from the beginning),
the response is strongly suppressed.

Here, we focus on yet another interesting phenomenon 
which our theory~\eqref{eq:TypTimeEvo}--\eqref{eq:ResProfEq} predicts
to occur
if the unperturbed system does {\em not} thermalize (e.g., due to the presence of conservation laws), 
but still equilibrates to its time-averaged value 
\begin{equation}
\label{eq:autav}
\Autav:=\lim_{T\to\infty} \frac{1}{T} \int_0^Tdt\, \< A \>_{\!\rho_0(t)}
\,.
\end{equation}
If the driving operator lifts this constraint,
Eq.~\eqref{eq:TypTimeEvo} can result in a three-stage relaxation process,
which can roughly be understood as a combination of ordinary (undriven) prethermalization and Floquet prethermalization ({\em ``Floquet double prethermalization''}).
The predicted effect is sketched in Fig.~\ref{fig:DoublePrethermSketch}.
An additional prerequisite is
that the time scales of $\< A \>_{\!\rho_0(t)}$, $\lvert \gamma(t, t) \rvert^2$, and eventual heating
are sufficiently well separated,
with the unperturbed relaxation $\< A \>_{\!\rho_0(t)}$ exhibiting the fastest time scale.
In this case,
we expect the following three stages:
(i)
The driven dynamics first closely resembles the undriven behavior and $\< A \>_{\!\rho(t)}$ approaches $\Autav$.
This is the ``ordinary prethermalization'' stage.
(ii)
As soon as 
$\lvert \gamma(t, t) \rvert^2$ starts to oscillate and deviate notably from 
unity (see below~(\ref{eq:ResProfEq})),
$\< A \>_{\!\rho(t)}$ moves on towards 
$\Auth$.
More precisely speaking, the last factor 
$\< A \>_{\!\rho_0(t)} - \Auth$
in ~\eqref{eq:TypTimeEvo}
approaches the constant value $\Autav-\Auth$, which is 
non-vanishing for suitable $A$ since the undriven system does not thermalize, 
while the second last factor
$|\gamma(t, t) \rvert^2$ still exhibits 
some (possibly damped) oscillations,
see also Sec.~\ref{sec:ResProf} and Ref.~\cite{dab24} for more details.
(iii)
Eventually, when heating sets in, $\< A \>_{\!\rho(t)}$ will tend towards $\Ainf$.

We remark that stage (ii) is akin to the previously known, so-called
``Floquet prethermalization'' effect \cite{kuw16, mor16,
aba17effective, aba17rigorous, mal19heating,
mac20, bea21, pen21},
which describes a plateau reached by the \emph{stroboscopic} 
dynamics encoded in the Floquet Hamiltonian.
Our present theory~\eqref{eq:TypTimeEvo} goes beyond this 
previous knowledge by predicting that,
due to the difference between $\Autav$ and $\Auth$,
the \emph{continuous} dynamics can show persistent oscillations during this stage.
This is the
first feature of the double-prethermalization effect
(which had already been discussed briefly in Ref.~\cite{dab24}).
The second feature is
stage~(i) and, in particular, the ``plateau'' at its end.
This feature will only be observable if the unperturbed 
dynamics $\< A \>_{\!\rho_0(t)}$ relaxes faster than the 
characteristic time scale of $\lvert\gamma(t, t)\rvert^2$ 
(typically the driving period $T$).
The final ``heating stage''~(iii), on the other hand,
is a commonly expected 
and observed
effect in generic driven many-body systems \cite{dal14, laz14equilibrium,
mal19heating, pon15manybody, ish18}
(however, it is not any more  captured by our theory~\eqref{eq:TypTimeEvo}).
Altogether, the occurrence of ``double prethermalization'' may thus not
appear to be entirely
surprising or novel in light of previous results.
Yet we believe it is worthwhile to note that stages~(i) and~(ii) of 
such a
behavior are captured quantitatively by 
Eq.~\eqref{eq:TypTimeEvo}.

To illustrate
the double-prethermalization effect, we consider as our third application another random-matrix model with a conservation law of $H_0$ that will be broken by the driving operator $V$.
Contrary to the random-matrix example from Sec.~\ref{sec:ResProf:Num} (Fig.~\ref{fig:ResProf}),
we choose the properties of the unperturbed Hamiltonian $H_0$, observable $A$, and initial state $\rho(0)$ in line with generic features of many-body systems.
The resulting model is arguably still less realistic than the spin model examples from Secs.~\ref{sec:FiniteQuench} and~\ref{sec:PseudrandomDriving},
but has the virtue that we can directly control all parameters entering the analytical prediction, similarly as in Sec.~\ref{sec:ResProf:Num},
meaning that no 
fitting of any 
parameter
will be 
necessary to compare numerics and theory.

The reference Hamiltonian $H_0 = \sum_\mu E_\mu \ketN{\mu} \braN{\mu}$
is defined via its $M = 2^{14} = 16\,384$ energy levels $E_0 := 0$ and 
$E_{\mu+1} := E_\mu + \varepsilon_\mu$ with $\varepsilon_\mu := \varepsilon_0 [ 1 + \alpha(1 + \cos(\frac{2\pi\mu}{M}))]$,
such that the density of states increases towards the middle of the spectrum [cf.\ Fig.~\ref{fig:ExampleModels}c1].
We choose $\alpha = 0.1$ and $\varepsilon_0$ such that the mean level spacing across the entire spectrum,
$\bar\varepsilon := (E_M - E_0)/M$, is $\bar\varepsilon = 2^{-9} = 1/512$ and hence $E_M = 32$.
The matrix elements $V_{\mu\nu}$ in the eigenbasis of $H_0$  satisfy the 
Hermiticity constraint ($V_{\mu\nu}^* = V_{\nu\mu}$)
and are drawn randomly from a complex ($\mu < \nu$) or real ($\mu = \nu$) Gaussian distribution with vanishing 
mean and variance $\varV(E_\mu - E_\nu) = \e^{- 2 |E_\mu - E_\nu|}$,
thus modeling a perturbation profile~\eqref{eq:VarV} of the exponentially 
decaying form \eqref{eq:VarV:Exp}, see also Fig.~\ref{fig:ExampleModels}c2.

To obtain an example for the above sketched ``Floquet double prethermalization'' scenario,
we assume that $H_0$ exhibits a ``conservation law'' that splits the spectrum into two sectors,
where states with even $\mu$ belong to the `$+$' and those with odd $\mu$ belong to the `$-$' sector.
The observable implements the ETH ansatz in the two sectors with 
\begin{eqnarray}
\label{eq:S:RMMPretherm:A}
A_{\mu\nu} := \delta_{\mu\nu} \, a_{\pm}(E_\mu) + R_{\mu\nu}
\ ,
\end{eqnarray}
where  $R = (R_{\mu\nu})$ is a random matrix from the Gaussian 
unitary ensemble (GUE with mean $0$ and variance $M^{-1}$) and
\begin{eqnarray}
a_\pm(E) := a^0_\pm \left[1 - \frac{2 (E - E_0)}{E_M - E_0}\right]
\ .
\end{eqnarray}
The infinite-temperature expectation value is thus $\Ainf = 0$ by construction.
Finally, $a^0_\pm$ are constants, whose values are choosen as
$a^0_+ = 1$ and $a^0_- = \frac{1}{4}$, see also 
Fig.~\ref{fig:ExampleModels}c3.

The initial state is of the form~\eqref{eq:InitState} with $Q = \id + \kappa A$ and $\ket\phi$ a Haar-random state in the `$+$' sector.
Similarly as in the examples from Figs.~\ref{fig:Spin12x2pQuench} and~\ref{fig:Pseudorand}, $\rho(0)$ thus mainly populates energy eigenstates within an
energy window of width $\Delta E = 4$ around $E = 12$, see also Fig.~\ref{fig:ExampleModels}c1.
In this window, the mean density of states [see above Eq.~\eqref{eq:Auth}]
is found to be
$D_0 \approx 500$ 
and the microcanonical expectation value 
$\Auth \simeq 0.15$.

Choosing the periodic step driving profile~\eqref{eq:DrivingProtocol:Step},
Fig.~\ref{fig:S:RMMPrepretherm} shows numerical 
results
(solid lines) of this system 
for various driving amplitudes $f_0$ and periods $T$ along with the theoretical prediction 
(dashed lines) obtained from Eqs.~\eqref{eq:TypTimeEvo}--\eqref{eq:ResProfEq}.
For clarity, the theoretical prediction is only shown for times $t \leq 10$
since it is not expected to apply for late times anyway.

We emphasize that the figure shows data for one particular realization of the entire 
model, i.e., no ``disorder averaging''
with respect to the randomness of 
$V$, $A$, 
and
$\ket\psi$ is performed.
Instead,
we recall that the theory includes a kind of self-averaging prediction,
namely that one particular realization of our random matrix model is 
expected to be 
very well approximated by the theory with very high probability 
\cite{dab24}.

According to Fig.~\ref{fig:S:RMMPrepretherm}, the prediction~\eqref{eq:TypTimeEvo}--\eqref{eq:ResProfEq} indeed
correctly describes the numerically observed behavior in the initial transient regime before heating becomes significant.
The accuracy is best for short times 
and remains reliable for longer times the smaller the driving period $T$
and/or the driving amplitude $f_0$ are.
Since the reference dynamics $\< A \>_{\!\rho_0(t)}$ is restricted to the `$+$' sector,
its long-time expectation value $\Autav = 0.25$ differs from the 
pertinent thermal equilibrium (microcanonical) value $\Auth = 0.15$ for 
the full system, see Eq.~\eqref{eq:Auth}.

Generally speaking, the various curves in Fig.~\ref{fig:S:RMMPrepretherm}
can be seen
to illustrate the above predicted three stages (i)-(iii).
In particular, oscillations between $\Autav$ and $\Auth$ [stage~(ii)] 
as well as the eventual approach of $\Ainf$ due to heating [stage~(iii)] 
can be directly observed in all examples.
The first initial plateau at $\Autav$ [stage~(i)] is best seen in 
Fig.~\ref{fig:S:RMMPrepretherm}b for $f_0 = 0.02$ (red curve) around $t \approx 1$,
albeit for a relatively short time only.
In the other examples,
it is certainly visible how the driven systems initially follow the 
unperturbed behavior $\< A \>_{\!\rho_0(t)}$ as (trivially) expected,
but
the latter does not decay sufficiently fast to be separated from the
onset of the oscillations encoded in $\lvert \gamma(t, t) \rvert^2$.
Unfortunately, we have not been able to achieve better separation 
of the time scales of $\< A \>_{\!\rho_0(t)}$ and $\lvert \gamma(t, t) \rvert^2$
within the numerically accessible system sizes for these examples.

More precisely speaking, the main difference between the idealized 
scenario in Fig.~\ref{fig:DoublePrethermSketch} and the numerical
data in  Fig.~\ref{fig:S:RMMPrepretherm} are the pseudo-random 
temporal fluctuations, which in turn would become weaker and 
weaker upon increasing the system size (in practice, they are 
seen to become stronger upon decreasing the system size).
In particular, for sufficiently large systems
the black curves in Fig.~\ref{fig:S:RMMPrepretherm}
are expected to become (practically) straight horizontal 
lines beyond $t\approx 0.5$.
Similarly, at least the red curves are expected to develop 
a ``nicer'' plateau  [stage~(i)] 
beyond $t\approx 0.5$.
We finally note that those temporal fluctuations 
in Fig.~\ref{fig:S:RMMPrepretherm} might seem to
be growing as $t$ increases,
but this is an artifact of the 
logarithmically plotted time axis.


\section{Conclusions}
\label{sec:Conclusions}

We revisited the recently developed typicality theory for the observable response of many-body quantum systems to time-dependent forcing from Ref.~\cite{dab24} and extended it in several directions:
First, we demonstrated its applicability for general, not necessarily periodic driving protocols in various different setups and found good agreement between the theoretical predictions and the observed dynamics in numerical experiments.
Second,
we derived analytical approximations for the function $\gamma(t, t')$ solving Eq.~\eqref{eq:ResProfEq},
which encodes the 
distinctly nonlinear
response characteristics in the main relation~\eqref{eq:TypTimeEvo},
in two limiting cases,
namely strong driving and fast driving.
Third,
we worked out implications of the theory in two physically interesting scenarios,
finite-time quenches and symmetry-breaking periodic driving.

It is important to point out that the theory~\eqref{eq:TypTimeEvo}--\eqref{eq:ResProfEq}
cannot be expected to hold universally for Hamiltonians of the form~\eqref{eq:H}.
At its core,
the relations~\eqref{eq:TypTimeEvo}--\eqref{eq:ResProfEq} are the solution of a random-matrix problem.
It asserts that, for a given unperturbed Hamiltonian $H_0$, initial state $\rho(0)$, and observable $A$,
the vast majority of driving operators $V$ will lead to observable dynamics as predicted by Eq.~\eqref{eq:TypTimeEvo}
for short-to-intermediate times and if heating effects are negligible.
This is clearly confirmed explicitly in Fig.~\ref{fig:ResProf}.

Similar models have been used to justify the eigenstate thermalization hypothesis (ETH) \cite{deu91,rei15, nat18, rei21, cip21, hel24},
to estimate error sensitivity in quantum simulators \cite{pog20},
and to predict the relaxation of
quantum systems under the influence of constant perturbations 
\cite{nat19, dab20relax, ric20, erd23}.
From a physical point of view,
the crucial question is whether the majority of members in the $V$ ensemble share those characteristics with the true driving operator of the system of interest that are responsible for the observable behavior.
Phenomenologically, it is clear that it is not necessary to know all microscopic details of a macroscopic system with, say, $10^{23}$ degrees of freedom.
Moreover, in equilibrium statistical mechanics it is a fundamental principle to use ensemble averages as predictions about an individual system after fixing some macroscopic properties such as energy, volume, particle number, etc.
In our present nonequilibrium setup,
the macroscopic property that is fixed and shared by the majority of driving operators $V$ is the coarse-grained energy profile $\varV(E)$ in the eigenbasis of $H_0$,
cf.\ Eq.~\eqref{eq:VarV}.
As demonstrated by the examples in Figs.~\ref{fig:Spin12x2pQuench} and~\ref{fig:Pseudorand} (see also Ref.~\cite{dab24} for further examples),
this captures the essence of the observable response
in some physically realistic setups.

Nevertheless, there are situations where
such a modeling may not be
sufficient or appropriate
\cite{rei16,ham18,rei19,nic20,hei21}.
An important example are cases where the dynamics involves, and the observable is sensitive to, 
macroscopic transport caused by the driving.
Since the $V$ ensembles do not take into account 
specific locality and geometrical properties
of the actual model system of interest,
situations where these properties matter macroscopically are likely to behave differently from
the ensemble-averaged prediction.
On the other hand, some locality information may be encoded in the unperturbed Hamiltonian 
$H_0$, whose induced dynamics
is assumed to be given in the theory.
As long as these properties are not changed substantially by the driving (as in the examples provided in Secs.~\ref{sec:FiniteQuench} and~\ref{sec:PseudrandomDriving}),
the theory may still be applicable.

Another class of relevant, but potentially not captured examples are cases where the observable $A$ in the true system of interest is strongly correlated with the driving operator $V$, e.g., if $A = V$ or $[A, V] = 0$.
Most of the operators $V$ in the ensembles considered here will not exhibit such a special relationship with the observable,
hence the resulting ``typical'' observable dynamics within the ensemble may differ noticeably from the behavior of the original physical system.
Devising suitable random matrix ensembles that can deal with such a situation is an interesting direction for future research.

\begin{acknowledgments}
This work was supported by the 
Deutsche Forschungsgemeinschaft (DFG, German Research Foundation)
under Grant No. 355031190 
within the Research Unit FOR 2692
and under Grant No. 
502254252.
This research utilized Queen Mary's Apocrita HPC facility, supported by QMUL Research-IT.
\end{acknowledgments}


\appendix

\section{Brief review of the derivation of Eq.~\eqref{eq:TypTimeEvo}}
\label{app:Derivation}

We briefly summarize key steps in the derivation of the response relation~\eqref{eq:TypTimeEvo} from Ref.~\cite{dab24},
focusing in particular on the aspects that are most relevant for 
our present explorations.

Given an initial state $\rho(0)$,
the state at any later time $\rho(t) = \mc U(t) \rho(0) \mc U(t)^\dagger$ can be expressed formally in terms of the solution $\mc U(t)$ of the Schr\"odinger/von Neumann equation $\frac{\d}{\d t} \mc U(t) = -\I H(t) \mc U(t)$,
cf.\ below Eq.~\eqref{eq:TimeEvo}.
We expand $\mc U(t)$ into a Magnus series \cite{bla09},
$\mc U(t) = \exp \sum_{k=1}^\infty \Omega_k(t)$,
where $\Omega_k(t)$ is a $k$-fold time integral over nested commutators of $H(t)$.
For typical many-body Hamiltonians,
guaranteed convergence of this series is restricted to very short times,
but it often continues to be useful as an asymptotic series \cite{bla09, mor16, kuw16}.
This means that finite truncations provide a useful approximation for significantly larger times,
where the reliability typically extends to longer times the shorter the characteristic time scale of variations of $H(t)$ is,
i.e., the faster the driving is \cite{bla09, buk15};
see, for example, Fig.~\ref{fig:ResProf} for a numerical illustration.
From a physical point of view,
it has been argued that the Magnus expansion breaks down when the energy absorption from the driving becomes significant \cite{dal13, dal14, ish18}.

We truncate the Magnus expansion at second order,
which leads to the approximation
\begin{equation}
	\mc U(t) \simeq \e^{-\I H^{(t)} t}
\end{equation}
with
\begin{align}
	H^{(t)} &:= H_0 + V^{(t)} \,,
	\\
	V^{(t)} &:= \frac{F_1(t)}{t} V + \left[ \frac{F_2(t)}{t} - \frac{F_1(t)}{2} \right] \, \I [V, H_0] \,,
\end{align}
where $F_1(t)$ and $F_2(t)$ were defined in~\eqref{eq:F}.

Next we consider the dynamics generated by the so-defined auxiliary Hamiltonians $H^{(t')}$
when starting from the same initial state $\rho(0)$ as our original system,
but treating $t'$ as a fixed parameter.
In other words, the time-evolved state of such an auxiliary system is $\rho(t, t') := \e^{-\I H^{(t')} t} \rho(0) \e^{\I H^{(t')} t}$,
where $t$ is the actual time and $t'$ is fixed.
Since
\begin{equation}
\label{eq:rhoAux}
	\rho(t) = \rho(t, t) \,,
\end{equation}
the auxiliary dynamics generated by $H^{(t')}$ can be used to extract the state of the true system of interest [under evolution with $H(t)$] at $t = t'$.

To predict the auxiliary dynamics,
we then adopt a typicality or random-matrix framework:
Instead of one specific driving operator $V$,
we introduce an ensemble of $V$ operators
that share the energy profile $\varV(E)$ from~\eqref{eq:VarV} with the true $V$ in the sense that $\av{ \lvert \braN{\mu} V \ketN{\nu} \rvert^2 } = \varV(E_\mu - E_\nu)$.
Here $\av{ \,\cdots }$ denotes the average over the $V$ ensemble and $E_\mu$ and $\ketN{\mu}$ are the eigenvalues and eigenvectors of $H_0$ [see above Eq.~\eqref{eq:Auth}].
The prediction for the dynamics of an individual $V$ is obtained in two steps:
First, 
it has been shown in  Ref.~\cite{dab24}
that the ensemble-averaged dynamics takes the form
\begin{equation}
\label{eq:AvgTimeEvo}
	\av{ \< A \>_{\!\rho(t, t')} }
		= \Auth + \lvert \gamma(t, t') \rvert^2 \left[ \< A \>_{\!\rho_0(t)} - \Auth \right] ,
\end{equation}
where
\begin{equation}
\label{eq:gammaFromU}
	\gamma(t, t') := \int \d E \, D_0 \, \e^{\I E t} \, u(E, t')
\end{equation}
is the Fourier transform of the function $u(E, t')$,
which characterizes the (ensemble-averaged) overlap of eigenvectors of $H^{(t')}$ and $H_0$,
\begin{equation}
\label{eq:u}
	\av{ \lvert \< n(t') \ketN{\mu} \rvert^2 } = u(E_n - E_\mu, t').
\end{equation}
Second, we 
established in Ref.~\cite{dab24}
that the probability to observe experimentally measurable deviations between $\< A \>_{\!\rho(t, t')}$ for a single, randomly chosen $V$ and $\av{\< A \>_{\!\rho(t, t')} }$
is exponentially suppressed in the number of degrees of freedom of the system.
For a macroscopic system with on the order of $10^{23}$ degrees of freedom,
this effectively turns the average~\eqref{eq:AvgTimeEvo} together
 with~\eqref{eq:rhoAux} 
 into the prediction~\eqref{eq:TypTimeEvo} for an individual realization of $V$.

The remaining task is to show that $\gamma(t, t')$ satisfies Eq.~\eqref{eq:ResProfEq}.
To this end, we exploit 
(see \cite{dab24} for details)
that $u(E, t')$ from~\eqref{eq:u}
can be expressed in terms of the ensemble-averaged resolvent
\begin{equation}
\label{eq:G}
	G(z - H_0, t') := \av{ (z - H^{(t')})^{-1} }
\end{equation}
[see above Eq.~\eqref{eq:ResProfFromG}] as
\begin{equation}
\label{eq:uFromG}
	u(E, t') = \frac{D_0}{\pi} \lim_{\eta\to 0+} \Im G(E - \I \eta, t') \,.
\end{equation}
The combination of Eqs.~\eqref{eq:gammaFromU} and~\eqref{eq:uFromG} provides the relation~\eqref{eq:ResProfFromG}.
By evaluating the ensemble average in~\eqref{eq:G} for large matrices $H^{(t')}$,
it can furthermore be shown that $G(z, t)$ satisfies Eq.~\eqref{eq:GAuxEq}.
Finally, the relation~\eqref{eq:ResProfEq} is found by multiplying both sides of Eq.~\eqref{eq:GAuxEq} by $\e^{\I x t}$,
integrating over $x \in \RR$,
and exploiting Eq.~\eqref{eq:ResProfFromG}.


\begin{thebibliography}{770}
\bibitem{fig:Spin12x2pQuench:FiniteSize}
The fact that the black curves in Fig.~\ref{fig:Spin12x2pQuench} are not exactly  
constant
is a finite-size effect of 
our dynamical typicality-like procedure to 
approximately imitate the initial thermal equilibrium 
state by a pure state according 
to (\ref{eq:InitStateQuench}).
Since the chains are isolated in the unperturbed system, the effective system size is only 14 spins.
In the fully interacting post-quench system
with 28 spins (colored curves in Fig.~\ref{fig:Spin12x2pQuench}), 
the finite-size effects are much weaker.
\end{thebibliography}
\end{document}